\documentclass[usenatbib]{mn2e}
\usepackage{graphicx,amsmath,amssymb}

\begin{document}
\topmargin-1cm

\title{Galaxy subgroups in galaxy clusters}

\author[J.D. Cohn]{J.D. Cohn\\
Space Sciences Laboratory, University of California, Berkeley, CA 94720}
\date{\today}

\pagerange{\pageref{firstpage}--\pageref{lastpage}} \pubyear{2011}

\maketitle

\label{firstpage}

\begin{abstract}
Galaxies which fall into clusters as part of the same infall halo can
retain correlations due to their shared origin.
N-body simulations are used to study properties of such 
galaxy subgroups within clusters, including their richnesses and prevalence.
The sizes, densities and velocity 
dispersions of all subgroups with
$\geq8$ galaxies are found and compared to those of the host clusters. 
The largest galaxy subgroup provides a preferred direction in the
cluster and is compared to other preferred directions in the cluster.
Scatter in cluster mass measurements (via five observables), along $\sim 96$
lines of sight, is compared to the relation of the line of sight to this preferred
direction: scatter in cluster velocity dispersion measurements show
the strongest correlation.  
The Dressler-Shectman test \citep{DS} is an observational method to
detect
cluster substructure.  This test is applied to the cluster sample
to see whether the substructure it identifies is related to these subgroups.
The results for any specific line of sight seem noisy, however,
clusters with large subgroups tend to have a higher fraction of
lines of sight where the test detects substructure.

\end{abstract}

\section{Introduction}
Galaxy clusters are of interest for many reasons (for some recent summaries and reviews see
e.g. \citet{Voi05,Kra09,Mye09,AllEvrMan11} and
the proceedings from the 2008 Enrico Fermi School on 
clusters\footnote{http://people.sissa.it/$\sim$lapi/efs08\_website/}).   They are the largest virialized objects
in the universe and as such extremely sensitive to the growth rate and
mass density of the universe.  They host unique astrophysical phenomena and the most
luminous galaxies in the universe and are more generally a special environment for transforming galaxies they enclose.
They were first discovered as clusters of galaxies (hence their name); but they
are also deep dark matter potential wells which gravitationally lens, and are
full of hot gas that emits in the X-ray and scatters cosmic microwave
background photons \citep{SunZel72}.   All of these different properties can be combined to
get a fuller understanding of the processes within and affecting galaxy clusters, and
their evolution and observable properties.  

Our interest here is in the galaxies in galaxy clusters, specifically those which joined the cluster as part of a shared infalling halo.  These will be called galaxy
subgroups hereon and are the focus of this note.
(Galaxies falling in from the same filament also are expected to have correlations
but will not be considered here.)  
Using N-body simulations with resolved subhalos\footnote{overdensities in phase space using the algorithm of \citet{DieKuhMad06}, see below} identified as
galaxies, we study statistical properties of these galaxy subgroups for over 200 clusters.   We measure their quantity, average properties, persistence, relation to 
some intrinsic cluster properties and mass observables, and how well
they can be detected via one classic observational test.

Some corollaries of the presence of these subgroups are already known \citep{KneGilGib04,Kne06,McI08,Angetal09,LiMo09,Sim09,WetCohWhi09,Kli10}.
For example, satellite-satellite
mergers within clusters and groups are not uncommon:
galaxies which are group members
when they fall in often merge with the center of their original infall group.

Groups of galaxies in clusters which are moving together or are more dense in
space are sometimes referred to as substructures.\footnote{The term substructure
can also refer to subhalos corresponding to the
galaxies themselves in simulations,
early work includes \citet{Kly99,MooGhiGov99}.}  X-ray cluster gas also
can exhibit substructure
(e.g. \citet{RicLoeTur92,Buo02} for early work and a review), as can the
cluster dark matter, e.g. mapped through gravitational
lensing (e.g. for a study on its effects on strong lensing of clusters see
\citet{HenDalBod07}).   These gas and dark matter substructures presumably have some relationship to the subgroups of galaxies of interest here.  Similarly, 
simulations are sometimes high enough resolution to 
resolve subhalos inside of subhalos
\citep{WelOstBod05,Sha07,Spr08,Gio09,LiMo09,YanMovan09,Gio10}.  If one identifies
these sub-subhalos as galaxies then a relation is implied between their
host subhalos and the galaxy subgroups described here.   Other related work includes the
tracking of groups of subhalos explicitly inside 8 simulated clusters \citep{Gil04}, and the measurements of subgroups in Milky Way size simulated halos, including their distribution and velocity 
associations \citep{LiHel08}. (Local 
group subgroups often 
can be identified using more information, such as metallicity
or three dimensional positional information.  There is a large body of literature
on the subject.)   Earlier observational studies of subgroups in
clusters in particular include analysis of ENACS (ESO Nearby Cluster Survey) clusters and Coma as well \citep{GurMaz98,GurMaz01}.
Previous work is extended here by the joint use of subhalos identified as galaxies (which preserves correlations between galaxies due to their histories), the subhalo and halo merging histories, and a large sample of clusters.
It expands upon the
subgroup properties and examples noted in
\citet{WCS} (WCS hereon), for the same simulation.

As galaxies moving together within a cluster can indicate the cluster is not relaxed
or, observationally, that interlopers are projected onto the cluster, 
tests have been designed to detect them.   
These tests (e.g. the Dressler-Shectman test \citep{DS})
have been applied to individual objects (including e.g. \citet{Bos06,Gir08,Bar11} )
larger cluster surveys (e.g. \citet{SolSalGon99,OegHil01,Bur04,HwaLee07,Ram07,Mil08,AguSan10,Ein10}) and numerical data (e.g. \citet{CroEvrRic96,Cen97,KneMue00,WCSl}).
An increase of galaxy and velocity substructure has been correlated \citep{EspPliRag07,RagPli07}
with more recent mergers,
higher density environments, and increased cluster elongation.
Pairs of galaxies in clusters were studied previously also, 
e.g. in \citet{denHar97,TayBab04}.

The work here is based upon N-body simulations, described in \S 2.  
Hydrodynamic simulations, which
include gas physics and a variety of subgrid prescriptions,
are not yet available at the volume and resolution considered here.
Some comparisons of subhalo properties with and without
hydrodynamics have been made \citep{Mac06,Dol08,Sar08,Sim09,JiaJinLin10,Kne10,SchMac11}.  In
some of these comparsions a small fraction
of radial orbits change (which are themselves a small fraction of subhalo orbits).
Direct application to the results here is not straightforward as the dark matter
subhalo finder used here is different than the ones for which the comparisons have
been made.

Simulations and mocks are described in \S 2, in \S 3  statistical properties of
the galaxy subgroups are given, \S 4 gives the relation of five mock observational mass measurements to
properties of galaxy subgroups and to each other, \S 5 describes the results of
applying the substructure
Dressler-Shectman test to these clusters, and \S 6 summarizes.

\section{Simulations}
A  $2048^3$ particle, $250 h^{-1} Mpc$ side periodic N-body 
simulation box is used, provided by
Martin White by running his TreePM \citep{TreePM} code. 
The cosmological parameters are
taken to be $h=0.7$, $n=0.95$, $\Omega_m=0.274$, and $\sigma_8=0.8$, in accord with a large number of cosmological observations.    The simulation has outputs at 45
times equally spaced in $\ln(a)$ from $z=10$ to $z=0$. 
Halos are found using friends-of-friends (FoF) \citep{DEFW} with a linking length of 
$b=0.168$ times the mean interparticle spacing. Halo masses given below are
FoF masses.  Our interest will be in the 243 clusters in the box, i.e. halos with $M\geq 10^{14} h^{-1} M_\odot$, at $z=0.1$.  (This redshift is used because some of the mock 
observations described below rely on models which were trained on observational data at this redshift.)

These simulation data were also used in WCS, which 
can be consulted for details of implementations, mock observations, and tests
beyond those given below.  Briefly,
galaxies are identified with subhalos and the two words will be used interchangeably
hereon.  The subhalos are found via the
FoF6d algorithm  of \citet{DieKuhMad06} (the specific implementation of their
algorithm
is described in the appendix of WCS). 
Subhalo infall masses can be used to infer galaxy luminosities  
(e.g. \citet{ConWecKra06}).  Here $\log_{10} M_{\rm inf} \geq 11.3$ is chosen, which 
corresponds to a minimum luminosity of
$\sim 0.2 L_*$ (at $z=0.1$ this is 
$\sim$ -18.5 in $r$ band \citep{Bla03} or  stellar mass 
$3 \times 10^9 h^{-1} M_\odot$ \citep{Mos10}).   No luminosity-infall mass scatter is
included (some estimates are in e.g. \citet{vdb07}).  In some cases as noted below,
$0.4 L_*$ is used as a minimum cut instead.  
This method of luminosity assignments for subhalos gives
agreement with observations for galaxy clustering, the cluster galaxy luminosity function, cluster richnesses and the radial cluster galaxy profile (see WCS).
To find galaxy histories, tracking is as described in \citet{WetCohWhi09,WCSl,WetWhi10}.

By augmenting the dark matter simulation, WCS measured mock observational 
masses for these clusters via several methods.  The relationship between the
observational masses and the simulated cluster masses, and the form of the scatter, are important for understanding the wealth of cluster survey data currently in hand
and arriving soon (see, e.g. the reviews cited above).
Five mass measurement methods from WCS are used here.   Two are richnesses. 
The first is the \citet{maxBCG} MaxBCG algorithm based
upon colors\footnote{MaxBCG is one of many algorithms based on a red
sequence finder \citep{RCSI,RCSII}.  Color assignments are estimated
with prescription of \citet{SkiShe09}, combined with redshift evolution of \citet{SP1,SP2,SP3}.
Galaxies are taken to be ``red'' if they have $g-r$ within 0.05 of the peak of the red
galaxy $g-r$ distribution specified by \citet{SkiShe09}, for their observed $M_r$, again see WCS for more detail.}.  The second richness estimator 
uses spectroscopy and assigns cluster
membership via the criteria of \citet{Yan07}).  The third mass measurement
uses Sunyaev-Zel'dovich (SZ) flux or Compton decrement.  Flux is assigned 
by using halo mass for temperature and taking the dark matter particle density proportional
to the gas density.  The flux is then
measured within an annulus of radius $r_{180b}$, the radius within which the
average mass is greater than or equal to 180 times background density.   Tests of
this approximation are given in \citet{WhiHerSpr02}. 
Weak lensing masses are found by using a cluster lens profile of SIS or NFW \citep{NFW} form
and then fitting for a velocity dispersion and then mass. The velocity
dispersion masses rely upon
phase space information
to reject outliers and include a measured harmonic radius in the mass
calculation, based on methods of 
\citet{vHaarlem97,denHartog, Kat96, Biv06, Woj07,Woj09}.\footnote{We will not use 
the velocity dispersions based upon 3 $\sigma$ clipping,
presented in WCS, as these were less well correlated with the FoF cluster mass.}  We use the WCS measurements corresponding to 
radius $r_{180b}$ when a radius needs to be specified. 
WCS measured each individual cluster's mass along 96 different lines of sight. 
This ensemble of cluster mass measurements will be used here as well, and
as in that analysis, lines of sight where a more massive cluster has its center 
within $r_{180b}$ are removed.   In addition, for the
work here, lines of sight where either richness was $<2$ were also discarded.

\section{Persistence of Galaxy Subgroups}
The inhomogeneities in cluster galaxy distributions are in part historical,
as larger halos are built up from the infall of smaller halos. 
When halos fall into larger halos, they become subhalos themselves, with their central and satellite galaxies all now becoming satellites within the new larger halo.   
To study these galaxy subgroups,
for every $z=0.1$ cluster galaxy\footnote{Only true cluster galaxies are included in this section, interlopers are considered in \S 5 and \S6 for observational mass estimates and substructure
estimates.}
 its infall group, 
infall time, infall group richness (above the luminosity cut) and infall group mass are identified.   Each infall halo containing more than one galaxy results in a
separate galaxy subgroup.
\begin{figure}
\begin{center}
\resizebox{3.5in}{!}{\includegraphics{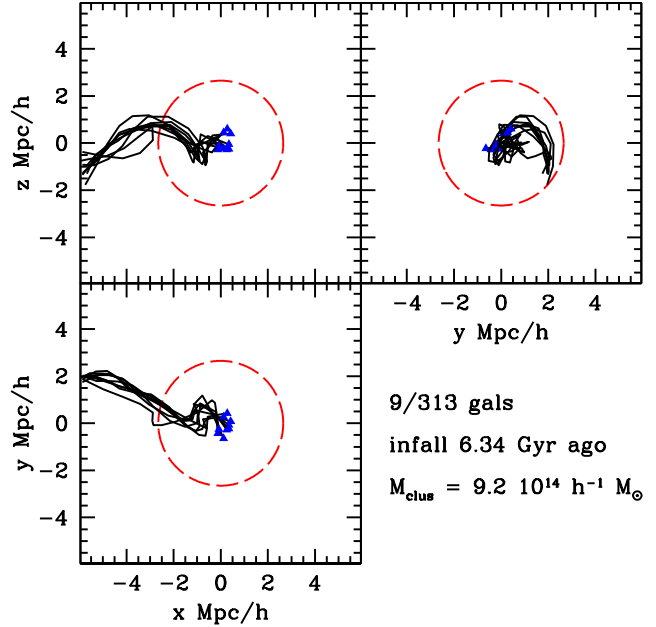}}
\end{center}
\caption{Tracks of subgroup galaxies after infall into a cluster.  
This subgroup fell into the cluster 6.3 Gyr
ago, and contained 12 galaxies at that time.   Only tracks of the galaxies surviving to $z=0.1$ are shown; solid triangles mark their $z=0.1$ positions.
The dashed circle is $r_{180b}$ for the full 9.2 $\times 10^{14} h^{-1} M_\odot$ cluster,
which has 313 galaxies.  The center position (0,0,0) is the cluster center at $z=0.1$.
}
\label{fig:tracks} 
\end{figure}
An example of a 9 galaxy subgroup is shown in Fig.~\ref{fig:tracks}, along with 
the tracks of these galaxies since their host halo fell into the
cluster 6.3 Gyr earlier (i.e. all the tracks are tracks of the subgroup members within the cluster itself).  At infall these
galaxies were in a $2.5 \times 10^{13} h^{-1} M_\odot$ halo with 12 galaxies.  Triangles
mark the final galaxy positions at $z=0.1$.  It can be seen that the galaxies have been
staying together throughout.  The dashed circle is $r_{180b}$ of the host
cluster at $z=0.1$, centered at the cluster center.
The host cluster has 313 galaxies and mass 9.2 $\times 10^{14} h^{-1} M_\odot$. 

A large fraction of galaxies within clusters are in these subgroups.  At our
redshift of interest,
$z\sim 0.1$, $>$40\% of the cluster galaxies
above $0.2L_*$  
have at least one associated companion \citep{WCSl}. 
(As mentioned earlier, these subgroup galaxies presumably correspond in some way to the sub-subhalos mentioned earlier \citep{WelOstBod05,Sha07,Spr08,Gio09,LiMo09,YanMovan09,Gio10}.)
Not all the galaxies which are not in subgroups fell in alone:
of the galaxies which have no companion in the clusters at $z=0.1$,
$\sim$ 1/5 fell in with companions which
have disappeared since infall.  
Some of the subgroup galaxies ($\sim 30\%$ of cluster galaxies) were preprocessed
in groups, i.e. fell in from halos with $M_{\rm halo} \geq 10^{13} h^{-1} M_\odot$.
The fraction of preprocessed cluster galaxies is similar to earlier measurements
\citep{Ber06,McGee09} (the simulated galaxy samples are not identical, so exact
agreement is not expected).  As in those papers,
a larger fraction of the more massive galaxies were processed in group
environments.
The importance of the group environment for galaxy formation has been
stressed in these papers and in e.g. \citet{Zab96,Zab07}, see also \citet{ZabMul98,ZabMul00,LiYeeEll09}.

\begin{figure}
\begin{center}
\resizebox{3.5in}{!}{\includegraphics{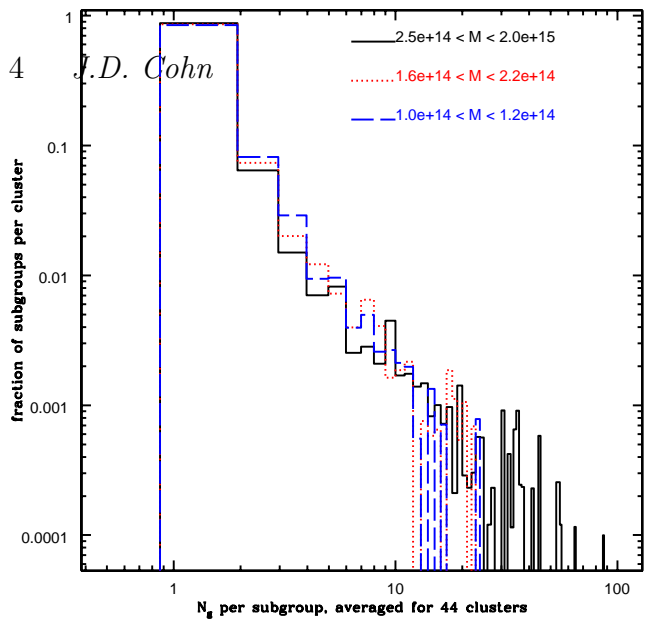}}
\end{center}
\caption{Fraction of subgroups, per cluster, with given richness, averaged over 44 clusters in three mass ranges.
The subgroup richnesses, similar to the
infall halo masses for clusters, seem to have a roughly self-similar distribution,
although there is a cutoff at high subgroup richness (i.e. $>20$) for lower mass clusters.}
\label{fig:subrich}
\end{figure}
The distribution of subgroup richness per cluster (i.e. average fraction of subgroups with given richnesses)
is shown in Fig.~\ref{fig:subrich} 
for three different mass regimes (44 clusters in each).
It appears to have a self-similar distribution, although there is a sharper
cutoff at high richness for the lower mass clusters.  (This is at
least in part
likely to be a resolution effect, as the maximum number of galaxies in
a subgroup is limited by the total number of galaxies in the cluster,
which is smaller for the lower mass clusters.)
There is likely some way to connect this subgroup distribution to the infall halo mass distribution,
such as measured in \citet{DeLuc04,Gao04,TayBab05,Gio09,Gao11}, 
as the infall group richnesses are related in infall masses (see also \citet{Sha07}
which compares instantaneous sub-subhalo masses).

\begin{figure}
\begin{center}
\resizebox{3.5in}{!}{\includegraphics{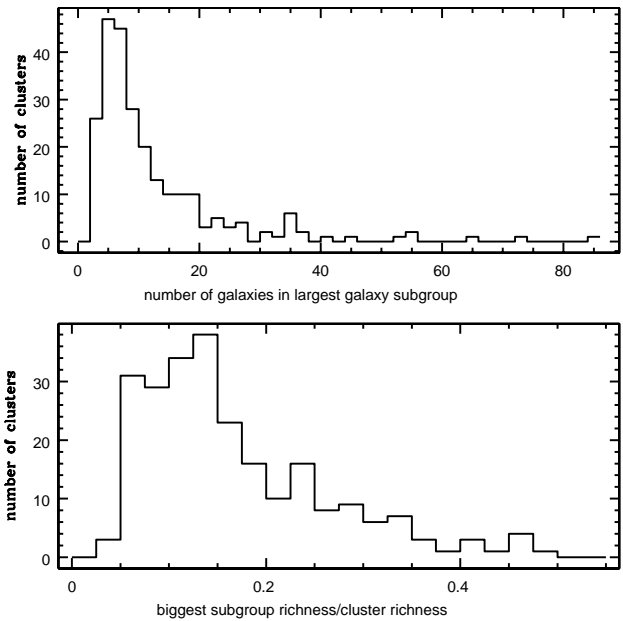}}
\end{center}
\caption{Top: for each of the 243 dark matter halos with
$M \geq 10^{14} h^{-1} M_\odot$, the largest number of galaxies remaining from the same original infall halo (subgroup richness). 
Bottom: the richness of the largest galaxy subgroup as fraction of its 
host cluster richness.   More massive halos are
are more likely to have more massive subgroups of galaxies.
The average value of largest subgroup richness over host cluster richness is 0.17. }
\label{fig:ngalsub}
\end{figure}

The richness distribution of the largest subgroup in each cluster is shown
in Fig.~\ref{fig:ngalsub}.  
Although the largest galaxy subgroup for many clusters has $\sim 10$ galaxies, much larger subgroups do occur.
The largest subgroup in our sample has
86 galaxies, and is in the second largest cluster ($M=1.1 \times 10^{15} h^{-1} M_\odot$).  This subgroup had 107 galaxies and host halo mass $2.4 \times 10^{14} h^{-1} M_\odot$
upon infall 2.4 Gyr ago.  Dividing largest subgroup richness by host cluster
richness, one sees that
the largest galaxy subgroup has on average about 20\% of the richness of its host
cluster, but that there is a wide scatter (Fig.~\ref{fig:ngalsub}, bottom). 
The ratio of subgroup richness to cluster richness might be expected to be
relevant because more massive halos tend to form 
from mergers of more massive progenitors, and richness is roughly proportional to mass.  (For
reference, a cluster of mass $10^{14} h^{-1} M_\odot$ has richness of $\sim 40$ on average.)
The particular richness fraction of the largest galaxy subgroup is not only determined by the infall mass of the cluster progenitor, but also the time since infall (i.e. the rate of disappearance of the subgroup galaxies and the growth of the cluster richness since infall).   Similarly,
the time since a recent major merger is correlated with the
size of the largest halo subgroup.  (Two standard criteria for a major merger
are that the smaller to larger mass is 1:3 or larger or 1:10 or
larger, both were considered here.)
 That is, the larger the subgroup the
more recent a major merger.  The strongest correlation is with the
fraction of the cluster richness due to the halo subgroup 
and the   
time of most recent 1:3 merger.  
(A related correlation was found by \citet{EspPliRag07,RagPli07}:
they found that measured cluster substructure increased for clusters
with recent mergers.)
The median subgroup richness fraction
(of the host cluster's richness) is about 1/3 for clusters with a
1:3 merger in the last time step (as required by the major merger definition), 
while for clusters with a 1:3 merger $\sim$ 10 Gyr ago,
the richness fraction is closer to 1/10.  This is not surprising as
after the major merger,
the cluster will grow in richness, while the subgroup will decrease.

\begin{figure} 
\begin{center}
\resizebox{3.5in}{!}{\includegraphics{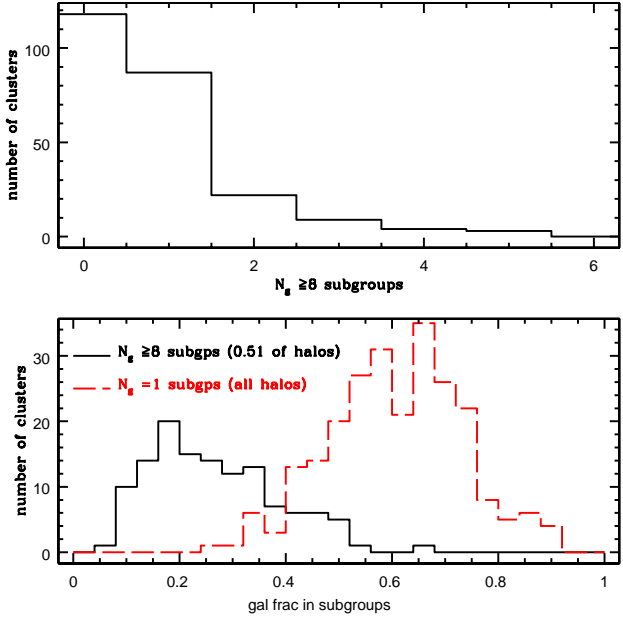}}
\end{center}
\caption{ Top: the number of large ($\geq 8$ galaxies) 
subgroups in all clusters. 
Bottom: (solid line) for the 50\% of clusters ($M\geq 10^{14} h^{-1} M_\odot$) with
subgroups of $\geq 8$ galaxies, the fraction of total cluster richness due to
these large subgroups, and (dashed line), for all clusters, 
the fraction of total cluster richness due to
galaxies with no associated galaxies in the cluster, per cluster.
 }
\label{fig:nclumpdist}
\end{figure}

Instead of the largest subgroup per cluster, one
can consider all ``large'' (defined to have $\geq 8$ galaxies, chosen for 
convenience\footnote{At $z=0.1$, richness
8 corresponds roughly to a halo with $M \sim 2 \times 10^{13} h^{-1} M_\odot$, but of course the subgroups are remnants of larger galaxy subgroups at infall. })  
galaxy subgroups. Their frequency per cluster is shown at the top of Fig.~\ref{fig:nclumpdist}:
118/243 clusters have no large subgroup, but
38 clusters have more than one large subgroup, and three clusters have 5 large subgroups. 
Almost all  (42 of the 44) clusters above $2.5 \times 10^{14} h^{-1}M_\odot$
contain at least one large subgroup.  
For lower mass host halos, large subgroups are more rare (e.g. occurring
only in 1/4 of the 725 halos with mass $\geq 5\times 10^{13}h^{-1} M_\odot$),
when they occur they are a larger fraction of the halo richness.  
Shown also in Fig.~\ref{fig:nclumpdist}, bottom, is the
fraction of cluster galaxy richness in large subgroups: the solid line is
richness fraction in large subgroups for the $\sim 50\%$ of clusters which contain
at least one large subgroup (peaked around 15-20\% but with a long tail to
large fractions).
For comparison, the richness fraction in all clusters due to
all galaxies not in subgroups is shown as well.  The latter is 60\%
on average,
as mentioned above, but has a wide spread.

\begin{figure}
\begin{center}
\resizebox{3.5in}{!}{\includegraphics{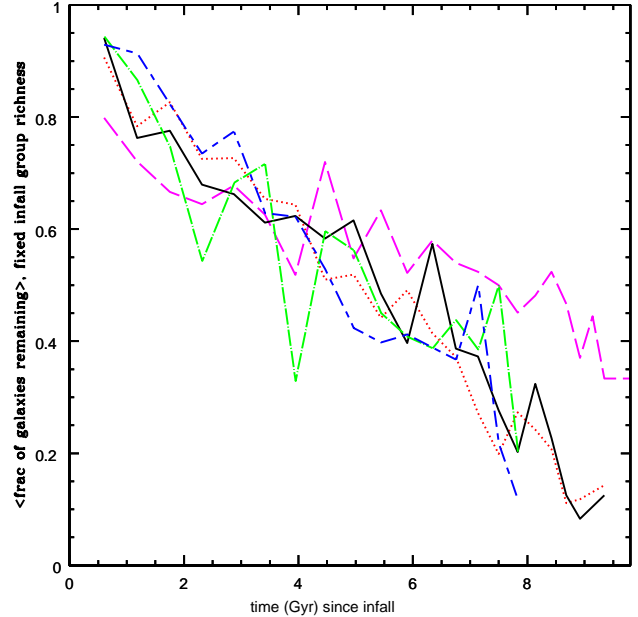}}
\end{center}
\caption{Average fraction of galaxies remaining for infall groups initially with (lines top
 to bottom at left):  $>$20 galaxies (dot-dashed), 15-20 galaxies (long-short dashed)
  11-14 galaxies (solid), 7-10 galaxies (dotted), and 3 galaxies (dashed),
in all halos with $M\geq 10^{14} h^{-1} M_\odot$ at $z=0.1$. Groups where all galaxies
 have disappeared are not shown.  Note the minimum possible fraction of galaxies
 remaining changes with infall richness.
}
\label{fig:fleft} 
\end{figure}

These $z=0.1$ subgroups are remnants of larger subgroups of galaxies upon infall.
Some subgroups fell in long ago: 
over half of the 189 large subgroups 
fell in more than 2 Gyrs earlier.
The subgroups which have fallen in more recently tend to have a larger
surviving fraction of galaxies. 
For example, the surviving 
fraction of galaxies per current large subgroup has a median value of $\sim$94\% for the 36 (currently) large subgroups whose halos fell in 0.6 Gyr ago, dropping to $\sim$ 50\% for those 26 falling in 4.9 or more Gyr ago.
Considering instead the {\it initial} richness of infalling groups,
the remaining fraction of galaxies (if nonzero) is shown for subgroups which have
3, 7-10, 11-14, 15-20, and $>20$ infall galaxies, as a function of
infall time, in Fig.~\ref{fig:fleft}.  Within the noise the change in galaxy fraction looks
similar for different initial galaxy richnesses (for subgroups starting with only 3
galaxies, the fraction is bounded below by 1/3).
Because of this galaxy attrition within subgroups, the satellite with the largest
infall mass (i.e. formerly the central galaxy of the largest infalling halo)
is not always in the largest subgroup.  A smaller halo
might have a larger remaining galaxy fraction.    For 70\% of
the  simulation clusters, the largest subgroup does
have the satellite with the largest $M_{\rm inf}$.
Another 5\% of the clusters have the satellite with largest
infall mass in a large subgroup (i.e. $\geq 8$ galaxies) but not the largest
subgroup.  When using subhalo
abundance matching (see, e.g. \citet{ConWecKra06}), the galaxy with
the largest infall mass will be the brightest satellite (up to
scatter).  However, given the above, even in the case of no-scatter
(i.e. oversimplified) halo abundance matching, the
brightest satellite will not be a useful way to identify the largest subgroup.

The spatial and velocity distributions of large subgroups differ on average from 
those of
their host clusters. 
As expected from their lower richnesses and their membership in a larger
cluster, subgroup sizes are generally smaller:  the average distance of subgroup members from the subgroup center is about half that of all cluster galaxies from the host cluster's center, albeit with scatter.\footnote{If the subgroup radii are rescaled by $M_{\rm inf}^{-1/3}$ of their infall halos,
and similarly for the full cluster and its mass, the ratio of subgroup to
cluster radii becomes very broadly centered around unity, that is, further suggesting that the smaller size of the galaxy subgroups is partially due to their smaller size at infall.  I thank M. George for suggesting
this and the rescaling below.}  
 (However, one of the 189 large subgroups has an average radius larger
than that of its host cluster.  
The subgroup center is taken to be the average
position, as not all subgroups have the infall central galaxy remaining.)
The average distance of subgroup galaxies from their 
center tends to be larger for older
subgroups.  One possibility is that galaxies close to the subgroup center 
have more time to merge with the subgroup center and disappear, while galaxies further away
have more time to be pulled away from the center by the host cluster tidal fields.

\begin{figure}
\begin{center}
\resizebox{3.5in}{!}{\includegraphics{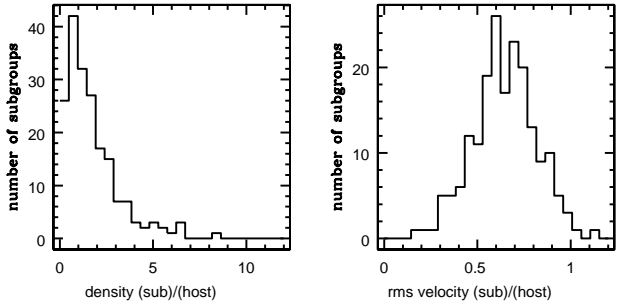}}
\end{center}
\caption{Left: The ``density''  (number of galaxies over
ellipsoidal volume) of large ($N_{g} \geq 8$) subgroups
divided by its counterpart for all galaxies in their host clusters.
 Right: velocity dispersion of galaxies
in large subgroups relative that of all galaxies in their host clusters.  For both left and
right, there can be more than one large subgroup in the same host cluster, as shown
in the top panel of  Fig.~\ref{fig:nclumpdist}.}
\label{fig:galden}
\end{figure}

In terms of density,
modeling the cluster and the subgroups as ellipsoids and then calculating
density as the number of galaxies divided by respective ellipsoidal volumes, 
the large galaxy subgroups
have a median (average) density $30\%(80\%)$ times larger than clusters they live in.  The broad distribution is shown at left in Fig.~\ref{fig:galden}--almost
1/3 of the large subgroups have densities which are more than double that
of their host cluster.  
In terms of shape, the median value of the (minimum axis/maximum axis)
for large subgroups divided by that for their host cluster is slightly below 1 ($\sim 4/5$), but there
is a tail out to values$>3$.  
That is, the majority of subgroups ($\sim 60\%$) tend to be less round 
than their host clusters.
Subgroups which fell in earlier tend to be more elongated
compared to those which fell in later (the average short/long axis ratio for
subgroups falling in at the last time step is $\sim$0.4, while for
subgroups which fell in over 6 Gyr ago, the average short/long axis
ratio is $\sim 1/4$).   
It might be expected that this elongation is due to tidal forces
within the larger cluster stretching out the subgroup after its infall.
However, there isn't a strong signal for alignment of the long axis of
the large subgroups with their direction of motion within the cluster.

In terms of velocities, the subgroups tend to have smaller ($\sim$70\%) velocity dispersions (relative to the average subgroup velocity) compared to their host cluster's counterpart, i.e. they are slightly colder.\footnote{Scaling the subgroup velocities by
their infall halo $M_{\rm inf}^{-1/3}$ and similarly for the cluster hosts actually gives
a median velocity dispersion which is higher for the subgroups, that is, the subgroups are colder than their hosts but hotter than isolated halos of similar mass.}
The ratio of velocity dispersions of large subgroups to those
of their host cluster  is shown in Fig.~\ref{fig:galden}. 
The subgroups which fell in earlier tend to have
larger velocity dispersions, both in raw numbers and (less strongly) relative to that of their
host halo. 

The radial distribution of the centers of the largest subgroup per cluster (rescaled
by the rms distance of the cluster galaxies from the cluster center) lies along
an NFW profile, just as the satellites in the simulation do \citep{WCSl} and as
has been seen observationally for cluster satellites in stacked profiles \citep{LinMohSta04}. 
It seems the outskirts of the stacked subgroup position profile are fit better by a lower concentration (i.e. 3), but the statistics are very noisy.

\begin{figure}
\begin{center}
\resizebox{3.5in}{!}{\includegraphics{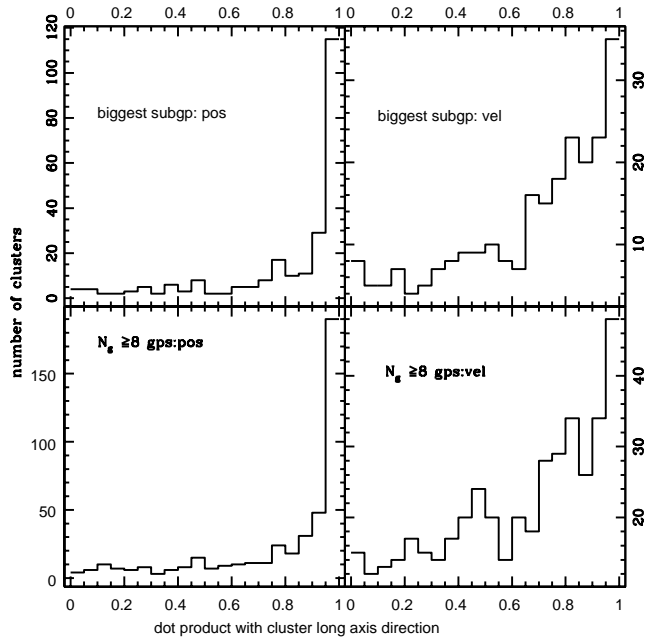}}
\end{center}
\caption{Left: The distribution of inner products of average position direction of the largest subgroup (top) and all large 
($\geq 8$ galaxies) subgroups (bottom) of galaxies, with the long axis direction of the
$M \geq 10^{14} h^{-1} M_\odot$ clusters, as determined by the moment of inertia of
its dark matter particles in its FoF halo.
Right: The distribution of inner products of the average velocity direction of galaxies within the largest subgroup (top) within the cluster and all large 
($\geq 8$ galaxies) subgroups (bottom) of galaxies, with the long axis direction of the
$M \geq 10^{14} h^{-1} M_\odot$ clusters.  The large ($\geq 8$ galaxies) subgroups and largest subgroup per cluster tend to
both lie along the long axis of the cluster and to be moving, in average, along
this axis.
}
\label{fig:pvdir}
\end{figure}

Clusters have several preferred directions, which often are related:
halos and thus their galaxies tend to fall in along filaments.
Filaments connecting clusters (i.e. usually the largest filament in
the cluster)
tend to lie along the cluster major axis, and the central halo is often
aligned with one or the other or both (references for cluster
alignments and formation along filaments include
\citet{vanBer96, Spl97, Col99, ChaMelMil00, OnuTho00, Fal02, van02, HopBahBod04, BaiSte05, Fal05, KasEvr05, LeeEvr07, Lee08, PerBryGil08, CosSodDur10}).
Alignments of cluster galaxy positions with each of these (and with a strength sometimes dependent upon color or morphology) has been seen in observations (e.g. \citet{Kne04,Bra05,Yan06,Azz07,Bai08,SteVal08,Wan08,Siv10,Wan10,Dea11,Lib11,Nie11}) and simulations (e.g. \citet{Col99,Ben05,KasEvr05,LeeKanJin05,Zen06,Lib05,AguBra06,Zen06,Kan07,KuhDieMad07,Lib06,Sal07,Fal08,Paz11}),
and has been characterized analytically as well (see, e.g. \citet{PenBen05,LeeKan06}).
As the largest subgroups tend to originate from massive halos, they might be expected to lie along the long axis of the cluster, i.e. the cluster's longest ellipsoidal axis. 
This is borne out in the simulated clusters: 
large subgroups and the largest subgroup per cluster
both tend to lie along the long axis of the cluster and to be moving (i.e. have average
velocity) along this axis (Fig.~\ref{fig:pvdir}). 
The clusters with longer large axes tend to have larger subgroups, and
subgroups further away from the center, however both of these may be
due to correlations between cluster mass and subgroup richness.  

Concentration is often strongly correlated with cluster history and other properties
(e.g. see \citet{Jee11,SkiMac11}).   In our sample, the cluster concentrations and deviations from the average concentration for a given mass were correlated
with several quantities (the size of the largest subgroup, the
deviation of the size of the largest subgroup from the average for that mass, 
the fractional richness in large subgroups, and the deviation of the fractional
subgroup richness in large subgroups from the average).  However,
these correlations were weak, $\sim$ 10\%-20\%.  They were in
the expected sense, that larger subgroups tended to be in clusters with smaller concentrations.

To summarize,
the majority of galaxies in clusters have no accompanying galaxies
remaining from their infall host halo, but galaxies in galaxy subgroups
still contribute a significant
fraction of cluster richness.  Galaxies in large subgroups ($\geq 8$ galaxies) tend to 
have smaller relative velocities and
higher relative densities than their hosts, even if their infall was several Gyr earlier.

\section{Subgroups and Masses}
Observationally, a cluster's galaxy population is often used to
find the cluster's mass, for instance via richness and velocity dispersions.
The large subgroups described in the previous section are deviations from
an idealized cluster of virialized galaxies within an isolated 
spherical dark matter potential.  As such,
they might affect the use of galaxies as tracers of cluster mass.
In this section the correlation between scatter in cluster mass measurements is
compared to properties of the large galaxy subgroups, when present.

In WCS, the dark matter simulations were augmented as 
described in \S 2, and mass measurements were made along 96 different lines 
of sight for every cluster, using 6 different techniques, 
for massive clusters ($M_{180b} \geq 2 \times 10^{14} h^{-1} M_\odot$).  In that paper,
some correlation between mass scatter and line of sight substructure (via the Dressler-Shechtman \citep{DS} test, see below) or large or small halos contributing 
galaxies outside the cluster was found.
In the same simulations \citet{NohCoh11} often found an increase in observable mass scatter when viewing along the plane containing the most (filamentary) mass in a 10 $h^{-1} Mpc$ radius sphere around
the cluster, or along the long axis of the cluster.
Substructure and observational mass scatter have been considered in other contexts, e.g.,
in X-ray measurements, \citet{Jel08} found mass scatter correlations with X-ray substructure.

It should be noted that since these mass measurements are only within the 250 $h^{-1} Mpc$ box, scatter due to the full line of sight is underestimated.  Both weak
lensing and Compton decrement measurements can easily have contributions to scatter
from well outside these scales.  This additional uncorrelated scatter would decrease the correlation of the observed scatter with intrinsic cluster quantities studied here.

Taking the largest subgroup of associated galaxies for each cluster, we
calculated the projection of its position on
viewing direction for $\sim$96 directions\footnote{Again, lines of sight with a more massive
cluster within $r_{180b}$ or with either richness $<2$ are discarded.}.  For many of the clusters, observed masses increased when measurements of
mass were along the largest subgroup's axis, i.e. looking along the direction where
the subgroup is almost directly behind or in front of the cluster center.

\begin{figure}
\begin{center}
\resizebox{3.5in}{!}{\includegraphics{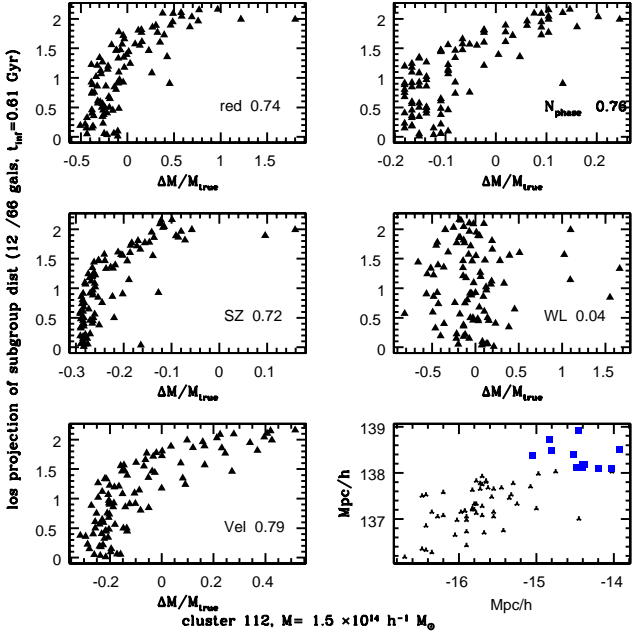}}
\end{center}
\caption{Correlation of mass scatter with projection of distance to
largest subgroup along line of sight, for a cluster with mass
$1.5 \times 10^{14} h^{-1} M_\odot$ and 66 galaxies, for (left to right, top to
bottom) red galaxy richness, richness via phase space, SZ, weak lensing and
velocity dispersions.  This cluster has 66 galaxies,
the largest subgroup at $z=0.1$ fell in 0.61 Gyr ago and has 12 galaxies and
the second largest subgroup has 6 galaxies and fell in almost 5 Gyr
ago. The center of the largest subgroup is 2.2 $h^{-1} Mpc$ from the cluster central galaxy.  Correlations
shown in each panel are calculated including all points shown.
At lower right, projected onto one plane, are the cluster galaxies
(pinwheels), with the largest subgroup galaxies represented by large filled in points.}
\label{fig:onecluscorr}
\end{figure}

One example in shown in Fig.~\ref{fig:onecluscorr}.  
This shows the fractional mass scatter for a cluster of mass
$1.5 \times 10^{14} h^{-1} M_\odot$ and 66 galaxies, for five mass
measurements.  Its largest subgroup fell in 0.61 Gyr ago and has
12 galaxies.  Most of
the observational mass estimates for this cluster increase as the line of sight tends to the
axis collinear with the largest subgroup (note this is also related
to the long axis of the cluster, as in Fig.~\ref{fig:pvdir}, and tends to be in the
plane containing most of the filamentary mass feeding the cluster).
Correlation coefficients are calculated using all points.
The correlation of scatter in mass is largest for the two measures of richness
and velocity dispersions, as might be expected.  (Again, the large SZ correlation
is likely overestimated, as is the correlation for weak lensing, because 
the small box size does not include uncorrelated scatter from larger scales
where lensing and SZ are also sensitive).  Also shown at lower right is a map of the cluster galaxies
with the largest subgroup galaxy positions noted (centered 2.2 $h^{-1} Mpc$ from the
central cluster galaxy).  Several of the clusters with recent mergers still have
the subgroups on the ``edge'' of the cluster, however, the dark matter density
in the region between these galaxies and the rest of the cluster galaxies is all above the threshold set by the $b=0.168$ linking length (roughly 100$\rho_b$).

The range of line of sight mass scatter varies by cluster and mass measurement.  
On a cluster by cluster basis,  the 65th-35th percentiles of mass scatter 
(i.e. the 65th - 35th percentiles of $(M_{\rm meas}-M_{\rm true})/M_{\rm true}$,
roughly the range of middle 1/3 of mass scatter) on average is 
$\sim$ 20\%, 10\%, 5\%, 30\% and 15\% for MaxBCG richness, phase richness,
SZ mass, velocity dispersion mass and weak lensing mass respectively.\footnote{This scatter is due only to changing the cluster line of sight.  
For all clusters, rather than one cluster along different lines of sight,
this width approximately doubles, except that for velocity dispersions, which only increases
slightly.  As SZ scatter is often closer to 20\% in larger boxes, e.g., \citet{CohWhi09},
its local scatter measured here is expected to double when uncorrelated structures
are included.} 
The maximum to minimum mass scatter per cluster due to line of sight is on average about 10 times this (for SZ it is $\sim \times 30$ ).
The width of the 65\%-35\% mass scatter per cluster increases with fraction of
richness in the cluster's largest subgroup for SZ and phase richness, and decreases
for weak lensing. 
The maximum to minimum widths are correlated with the fraction of cluster richness in the largest subgroup when the subgroups are at least 20\% of the
cluster's richness (69/243 of the clusters).

\begin{figure}
\begin{center}
\resizebox{3.5in}{!}{\includegraphics{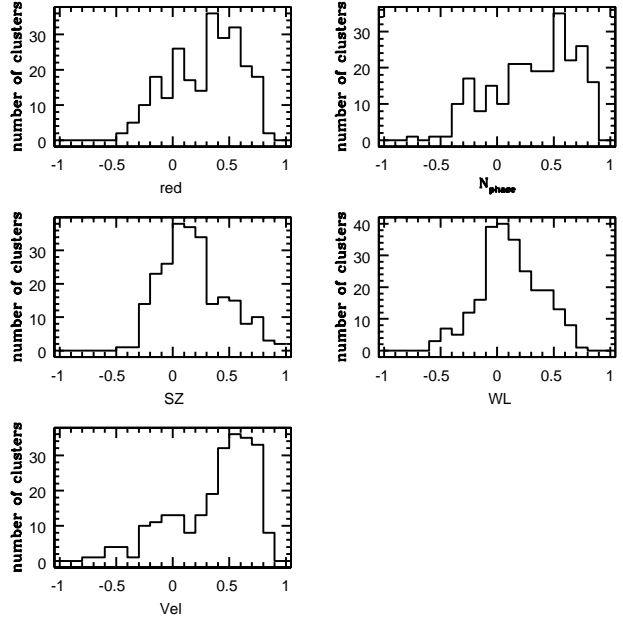}}
\end{center}
\caption{For each of the 243 dark matter halos with
$M \geq 10^{14} h^{-1} M_\odot$, the correlation between
the distance to the largest associated subgroup of galaxies, projected
on the line of sight, and the mass scatter.  Measuring mass along the direction of the
largest subgroup of galaxies often produces increased scatter in masses based upon
 richness and velocity dispersions.  Smaller correlations are seen for SZ
and weak lensing (it should be noted that for these two measurements that
the simulation box size is relatively small compared to
the line of sight distances which contribute to their scatters).}
\label{fig:galanglecorr}
\end{figure}

For these five different observable masses, one can again measure the distribution of
correlations between observational mass scatter and projection on the
direction of the largest subgroup of the line of sight.  These correlation 
coefficients are shown in
Fig.~\ref{fig:galanglecorr}.
Although not all clusters had correlations between the line of sight to the largest
subgroup and mass scatter, several did.
Richness and velocity dispersions tend to be enhanced when the line of sight
tended to the direction of the largest subgroup in the cluster, as in the
example in Fig.~\ref{fig:onecluscorr}.  These trends can be 
compared to the correlations of mass scatter
with the long axis of the cluster and with measurements through or within the
filamentary plane around the cluster,  studied in
\citet{WCSl,NohCoh11}.  
Correlations of mass scatter with respect to the direction
of the largest subgroup are comparable but smaller for the two richness measurements
and much smaller for SZ and weak lensing.  However,
the average correlation between mass scatter and direction of observation relative
to the largest subgroup is
much larger for velocity dispersions than the correlations found using other cluster preferred axes.  

Combining all the clusters, the correlation between
mass scatter and direction to largest subgroup, or largest subgroup if one is
present with $\geq 8$ galaxies, is much weaker.  This is presumably due
to the tails into negative correlations which are seen for some
clusters in Fig.~\ref{fig:galanglecorr}.  One possible cause for weak or opposite
sign correlations between mass scatter and the line of sight for measurement
versus direction to largest subgroup is
the presence of more than one large subgroup in many clusters (e.g. Fig.~\ref{fig:nclumpdist}).  

Another quantity was explored as well,
the correlation of scatter with projection onto the position of
the satellite galaxy with largest $M_{\rm inf}$.  This galaxy would be, in the
absence of luminosity-infall mass scatter, the brightest non-central galaxy. 
The relation between the direction to this galaxy and the line of sight of mass
measurement was not as strongly correlated with the mass scatter (i.e. the distribution
of correlation coefficients for all the clusters tended to be centered on zero).
\footnote{Many of the mass scatters are correlated with each other,
due to their origin in properties in and around the cluster, and have been
studied in e.g.
\citet{Nor08,CohWhi09,Men10,Sta10,WCSl}.
As discussed in \citet{Nor08,Ryk08, Sta10,WCSl},  correlated mass scatter
can not only cause underestimation of mass errors for single objects, but
can also introduce biases when measurements are stacked.}

To summarize, mass scatters in cluster measurements, for several different methods, are correlated with relative directions between the line of sight and the axis pointing towards the largest cluster galaxy subgroup, similar to correlations found earlier between mass scatter and directions of observation relative
to the long axis of the cluster and to the filamentary plane around the cluster. 
Velocity dispersion mass scatter is more correlated with the direction to the largest
galaxy subgroup than with the long axis or filament plane directions.

\section{Finding subgroups observationally}
As seen in earlier sections, large galaxy subgroups are frequent
in massive clusters, and the relation of their position relative to the line
of sight often correlates with mass scatter, particularly velocity dispersions.  
The ability to detect these
subgroups of associated galaxies thus may be useful both for their own interest
and for studies aimed at galaxy cluster masses.  The presence or
absence of galaxy substructures (galaxies close in space with smaller relative
velocities) is also sometimes
used to estimate whether a cluster
is relaxed or not, for example to allow the assumption of hydrostatic equilibrium
for X-ray analyses.   
Several techniques exist for finding galaxy substructure within clusters (e.g.
\citet{DS,Pin96,KneMue00,GurMaz01,Hou09} and those noted earlier).  These have been used
in several observational and numerical studies including \citet{CroEvrRic96,Cen97,SolSalGon99,KneMue00,OegHil01,Bur04,Bos06,HwaLee07,Ram07,Gir08,Mil08,AguSan10,Ein10,WCSl,Bar11}.

Here we apply the classic test by 
\citet{DS}\footnote{\citet{Pin96} find that DS is the most sensitive to substructure,
but that it also has the highest detection of non-substructures due to elongation and
velocity dispersion gradients.} to see how its substructure detection
criterion
relates to the presence of large subgroups.
The study here differs from previous work by focussing on a specific sort of substructure present in the clusters, substructure in galaxies coming from the
same infall group.   It also differs
from some previous works as it uses subhalos rather than dark matter particles
as galaxies in applying the test, and thus can apply interloper methods to
identify cluster galaxies in closer
analogy to observations.  (That is, just as in WCS but not in most previous
work, whole subhalos, with dark matter central position and mean
velocity, are used as galaxies for the tests, rather than random dark matter particles.)

The original DS test checked for collections of $N_{nn}+1=11$ galaxies whose relative velocities are less likely than random within the cluster (``random'' is calculated
by shuffling the velocities of the galaxies in the cluster).  
One calculates, centering on each galaxy,
\begin{equation}
\delta_{gal}^2 = \frac{N_{nn}}{\sigma^2}\left[(\bar{v}_{\rm local} -\bar{v})^2 +(\sigma_{\rm local} - \sigma)^2 \right] \; .
\end{equation}
In the rest frame of the cluster, $\bar{v}$ and $\sigma$ are the 
mean velocity and velocity dispersion of all the cluster galaxies, while the local 
counterparts include only the $N_{nn}$ nearest galaxies to the chosen galaxy, 
in the plane of the sky. 
The sum $\Delta = \sum_{gal} \delta_{gal}$ over all cluster galaxies is used to estimate
whether substructure is present.  In the original formulation, $\Delta$ is compared
with random velocity shufflings in the same cluster.
A low fraction of shuffled directions giving $\Delta$ larger than
the unshuffled $\Delta$ ($P \leq 0.05$) is considered a substructure detection.
\citet{Pin96} take instead $N_{nn}$ as the
square root of the total number of cluster galaxies and \citet{KneMue00}
take $N_{nn} = 25$.  The latter found that $\Delta/N_{\rm gal} \geq 1.4$ was also a 
useful criterion to detect substructure, using dark matter particles, other
studies have found that a mass dependent threshold in $\Delta$ is more
accurate \citep{RagPli07}.  In addition to choosing the number of nearest
neighbors, $N_{nn}$, it is also necessary to specify a luminosity threshold when
applying the test, and to select which galaxies are in the
cluster.

For application to our mock catalogues, 
a priori knowledge of true cluster galaxies cannot be assumed if observations
are to be mimicked.  Cluster
membership for galaxies must instead
be determined using observational techniques.  The original DS test used $3-\sigma$
clipping, but more sophisticated methods exist for identifying cluster members
(for example, \citet{vHaarlem97,denHartog,Biv06,Woj07,Woj09}) which take into account both the line of sight velocities
and the distance from the estimated cluster center.  We use the variant of these
methods as outlined
in WCS to identify cluster galaxies.   
\begin{figure}
\begin{center}
\resizebox{3.5in}{!}{\includegraphics{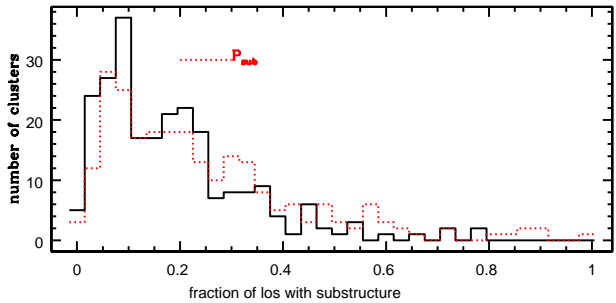}}
\end{center}
\caption{Fractions of lines of sight, per cluster, which have substructure 
by the  Dressler-Shectman test (described in text).  The number of galaxies
per test subgroup is taken to be
the square root of the number of cluster galaxies (found via method described in
text as well) with $L\geq 0.4 L_*$.
The solid line corresponds to the number of clusters where substructure is detected
along a given fraction of sightlines, via the condition $\Delta/N_{\rm
  gal} \geq 1.4$ (in principle only calibrated for $N_{nn}=25$). 
The dotted line corresponds to the number of clusters where substructure is detected
along a given fraction of sightlines by
the requirement $P_{sub} \leq 0.05$.   
As the number of galaxies
per test subgroup is changed from 10 or 25, two other canonical values used
in the literature, the fraction of sightlines per cluster with substructure decreases.
The case shown here tends to have more substructure than these other two choices,
because most of the clusters have the square root of number of galaxies $<10$,
and substructure detections seem to decrease as $N_{nn}$ increases.
 }
\label{fig:dsdist}
\end{figure}
We considered $M_{\rm inf}$ cuts corresponding to $L\geq 0.2, 0.4 L_*$,  
and
$N_{nn} = \sqrt{N_{g}}, 10, 25, N_{g}/6$.  This last was aimed at getting
at the natural subgroup size implied by Fig.~\ref{fig:ngalsub}.

The presence of cluster substructure seemed to depend strongly upon viewing angle.
Although for some clusters substructure was detected more often
when the line of sight was perpendicular to the largest subgroup, for
other clusters no correlation was apparent between the direction to
the largest subgroup and the amount of substructure found.
Taking in principle the cleanest cases, the 87 clusters with only one 1 large subgroup, substructure tended to be
more often detected if the subgroup was perpendicular to the line of 
sight.  In this case the subgroup was on the 'side' of the cluster in the plane of
the sky.  But even here, $\sim 1/4$
of the clusters had substructure detection increasing as the axis to the substructure became more closely aligned to the line of sight.

Rather than considering substructure detection along any particular line of
sight, one can instead consider the number of lines of sight with substructure, as
a property of a given cluster.
Fig.~\ref{fig:dsdist} shows the distribution, for all the clusters, of the 
fraction of lines of sight where substructure
is found, by both measures, using $N_{nn}=\sqrt{N_{g}}$, and
minimum infall mass $M_{\rm inf} \geq 11.69$ (to roughly correspond to
$L\geq 0.4 L_*$).  The dotted line corresponds to the fraction of lines
of sight which meet the substructure detection requirement of
$P_{\rm sub} \leq 0.05$. The numbers of lines of
sight with substructure with the two measures tended to be correlated with each other.
Hardly any clusters had no substructure along any line of sight.  Many
clusters had 10\% of their lines of sight showing substructure, and a handful had
over half of their lines of sight showing substructure.

The results depended upon the parameters and substructure criterion.  Substructure detection using the 
requirement of $P_{sub}\leq 0.05$ occurred more often than
detection using the requirement $\Delta/N_{nn} \geq 1.4$ (and almost always detected
substructure when
the latter test did).  
As the number of nearest neighbor galaxies used in the finder increased
(i.e. larger substructures were sought), substructure was found along fewer lines
of sight.  Conversely, as the luminosity cut was lowered to include more
galaxies, substructure was found more often.  Lines of sight with substructure
were not in 1-1 correspondence for different luminosity cuts, but the fraction of
lines of sight with substructure was correlated cluster by cluster as
the luminosity cut changed. 

The number of lines of sight with substructure per cluster was correlated
with the number of galaxies in the largest substructure, and the number of
galaxies in subgroups with $\geq 8$ galaxies. 
(This is likely also related to the increase of substructure found for
recent mergers noted in \citet{EspPliRag07,RagPli07} and mentioned earlier.)
More substructure was found in higher mass clusters as well.  To try to
minimize the mass dependence, clusters were divided into 6 mass bins and
substructure lines of sight were counted for both 
the top and bottom quartiles of galaxies in large subgroups.  
In these bins the clusters with more galaxies in large subgroups tended to
have more lines of
sight with substructure (but not always, and the results varied with substructure
test criteria).
If only subgroups which had fallen in recently (i.e. within 0.6 Gyr) were considered,
then substructure was more often detected for clusters whose largest subgroup
was farthest from the cluster center in three dimensions (with a stronger correlation with increasing
$N_{nn}$ in the substructure test). 

In summary, the Dressler-Shectman test did tend to detect substructure more
often in clusters with larger subgroups, but the results varied strongly between
lines of sight.   The increased frequency of sightlines which detect subgroups as
the subgroup size increases is encouraging.  However, the lack of
substructure as defined by this test does not necessarily mean the substructure is
absent.  This suggests using caution when taking the results of the DS
test to measure whether substructure is present or whether the
cluster is in hydrodynamic equilibrium (e.g., for the application of X-ray
mass estimates).


\section{Summary}
This note considered subgroups in clusters, i.e. galaxies which shared the same halo
before they fell into the galaxy cluster, in a relatively
large volume and high resolution N-body simulation.
The cluster sample had
243 clusters at $z=0.1$.   Properties of the distribution of subgroup populations, and how their orientations relative to
line of sight of observation affected a variety of cluster mass measurements were considered,
along with one test of cluster substructure.

 A significant fraction of cluster galaxies are in these subgroups, 
with larger subgroups more likely in more massive (i.e. richer) clusters.
In roughly half of the simulation clusters with $M \geq 10^{14} h^{-1} M_\odot$, 
at least one large ($\geq 8$ galaxies) subgroup is present, and 15\% of the
clusters have more than one large subgroup.  The large subgroups tend to have
higher galaxy densities and smaller rms velocities than the
full sample of galaxies in their host cluster.  These correlations often remain
many Gyr after infall.
The largest subgroups tend to be found along the cluster long axis, with
average velocities tending to be directed along this axis as well.
Observationally, cluster mass measurements on average increase when
the largest cluster subgroup lies along the line of sight,
cluster by cluster, for two richness mass measurements
and especially for velocity dispersions.
The larger the cluster subgroup, the more likely substructure will be detected by
the Dressler-Shectman test, but the likelihood of detection also depends upon
the line of sight direction, the number of nearest neighbors used in the test, and the luminosity cut for the galaxies included.

It would be interesting to see how these subgroups evolve compared to the
gas of their original infall halos, and to understand what other galaxy properties
(besides mergers and those discussed in \citet{KneGilGib04,Kne06,McI08,Angetal09,LiMo09,Sim09,WetCohWhi09,Kli10})
depend upon subgroup membership for these cluster galaxies.  

\section*{Acknowledgements}

JDC thanks M. George, S. Ho, A. Leauthaud and Y. Noh for discussions, and especially
thanks M. White for numerous discussions, and
for use of his simulation and mock catalogue data.   She thanks M. George, A. Knebe and
M. White for suggestions on the draft as well.  Last, but not least,
she also thanks the anonymous
referee for several very useful suggestions and comments.
The simulations used in this paper were performed at the National Energy
Research Scientific Computing Center and the Laboratory Research Computing
project at Lawrence Berkeley National Laboratory, analysis was also done
on the computer funded by the 2009 NSF ATI:{\sl Acquisition of a Beowulf Cluster for Computational Astrophysics, Cosmology and Planetary Science at
UC Berkeley}.

\label{lastpage}

\end{document}